\documentclass[epj,a4paper]{svjour}
\usepackage{multirow,rotating}
\usepackage[latin1]{inputenc}  
\usepackage{graphicx}

\graphicspath{{../Figures-EPS/}{.}{}}
\usepackage[]{color}
\usepackage[]{graphicx}
\usepackage[]{enumerate}
\usepackage[]{lineno}
\usepackage[square,comma,numbers,sort&compress]{natbib} 

\newcommand{\infnge}{1}
\newcommand{\infnpd}{2}
\newcommand{\unipd}{3}
\newcommand{\hzdr}{4}
\newcommand{\tudresden}{5}
\newcommand{\edinburgh}{6}
\newcommand{\infnna}{7}
\newcommand{\atomki}{8}
\newcommand{\lngs}{9}
\newcommand{\infnto}{10}
\newcommand{\mi}{11}
\newcommand{\roma}{12}
\newcommand{\teramo}{13}
\newcommand{\bochum}{14}

\begin{document}

\title{A new study of the $^{22}$Ne(p,$\gamma$)$^{23}$Na reaction deep underground: Feasibility, setup, and first observation of the 186 keV resonance}
\author{
	F.\,Cavanna\inst{\infnge} \and
	R.\,Depalo\inst{\infnpd,\unipd} \and
	M.-L.\,Menzel\inst{\hzdr,\tudresden} \and
	M.\,Aliotta \inst{\edinburgh} \and
	M.\,Anders\inst{\hzdr,\tudresden} 
	D.\,Bemmerer \inst{\hzdr}  \thanks{d.bemmerer@hzdr.de} \and
	C.\,Broggini \inst{\infnpd} \and
	C.G.\,Bruno \inst{\edinburgh} \and
	A.\,Caciolli \inst{\unipd,\infnpd} \and
	P.\,Corvisiero \inst{\infnge} \and
	T.\,Davinson \inst{\edinburgh} \and
	A.\,di Leva \inst{\infnna} \and
	Z.\,Elekes \inst{\atomki} \and
	F.\,Ferraro\inst{\infnge}
	A.\,Formicola \inst{\lngs} \and
	Zs.\,F\"ul\"op \inst{\atomki} \and
	G.\,Gervino \inst{\infnto} \and
	A.\,Guglielmetti \inst{\mi} \and
	C.\,Gustavino \inst{\roma} \and
	Gy.\,Gy\"urky \inst{\atomki} \and
	G.\,Imbriani \inst{\infnna} \and
	M.\,Junker \inst{\lngs} \and
	R.\,Menegazzo \inst{\infnpd} \and
	P.\,Prati \inst{\infnge} \and
	C.\,Rossi Alvarez \inst{\infnpd} \and
	D.A.\,Scott \inst{\edinburgh} \and
	E.\,Somorjai \inst{\atomki} \and
	O.\,Straniero \inst{\teramo} \and
	F.\,Strieder \inst{\bochum} \and
	T.\,Sz\"ucs \inst{\hzdr} \and
	D.\,Trezzi \inst{\mi} 
	(LUNA collaboration)
	}                     
\institute{
	Dipartimento di Fisica, Universit\`a di Genova, and Istituto Nazionale di Fisica Nucleare (INFN), Sezione di Genova, Italy 
	\and
	INFN Sezione di Padova, Padova, Italy 
	\and 
	Dipartimento di Fisica e Astronomia, Universit\`a di Padova, Padova, Italy 
	\and
	Helmholtz-Zentrum Dresden-Rossendorf (HZDR), Dresden, Germany 
	\and
	Technische Universit\"at Dresden, Dresden, Germany 
	\and
	SUPA, School of Physics and Astronomy, University of Edinburgh, Edinburgh, United Kingdom 
	\and
	Dipartimento di Fisica, Universit\`a degli Studi di Napoli Federico II, and INFN Sezione di Napoli, Napoli, Italy  
	\and 
	Institute of Nuclear Research of the Hungarian Academy of Sciences (MTA ATOMKI), Debrecen, Hungary 
	\and
	INFN, Laboratori Nazionali del Gran Sasso, Assergi, Italy 
	\and
	Dipartimento di Fisica Sperimentale, Universit\`a di Torino, and INFN Sezione di Torino, Torino, Italy 
	\and 
	Universit\`a degli Studi di Milano, and INFN Sezione di Milano, Milano, Italy 
	\and
	INFN Sezione di Roma ``La Sapienza'', Roma, Italy, 
	\and
	Osservatorio Astronomico di Collurania, Teramo, and INFN Sezione di Napoli, Napoli, Italy 
	\and
	Ruhr-Universit\"at Bochum, Bochum, Germany 
}
\date{\today}

\abstract{
The $^{22}$Ne(p,$\gamma$)$^{23}$Na reaction takes part in the neon-sodium cycle of hydrogen burning. This cycle is active in asymptotic giant branch stars as well as in novae and contributes to the nucleosythesis of neon and sodium  isotopes. In order to reduce the uncertainties in the predicted nucleosynthesis yields, new experimental efforts to measure the $^{22}$Ne(p,$\gamma$)$^{23}$Na cross section directly at the astrophysically relevant energies are needed. In the present work, a feasibility study for a $^{22}$Ne(p,$\gamma$)$^{23}$Na experiment at the Laboratory for Underground Nuclear Astrophysics (LUNA) 400\,kV accelerator deep underground in the Gran Sasso laboratory, Italy, is reported. The ion beam induced $\gamma$-ray background has been studied. The feasibility study led to the first observation of the $E_{\rm p}$ = 186\,keV resonance in a direct experiment. An experimental lower limit of 0.12\,$\times$\,10$^{-6}$\,eV has been obtained for the resonance strength. Informed by the feasibility study, a dedicated experimental setup for the $^{22}$Ne(p,$\gamma$)$^{23}$Na experiment has been developed. The new setup has been characterized by a study of the temperature and pressure profiles. The beam heating effect that reduces the effective neon gas density due to the heating by the incident proton beam has been studied using the resonance scan technique, and the size of this effect has been determined for a neon gas target.
\PACS{
	{26.30.-k}{Nucleosynthesis in novae, supernovae, and other explosive stars} \and
	{25.40.Ep}{Inelastic proton scattering} \and
	{29.30.Kv}{X- and gamma-ray spectroscopy} \and
	{51.20.+d}{Viscosity, diffusion, and thermal conductivity}
    }
}
\authorrunning{F. Cavanna {\it et al.} (LUNA collab.)}
\titlerunning{A new study of the $^{22}$Ne(p,$\gamma$)$^{23}$Na reaction deep underground...}

\maketitle

\section{Introduction}

The observed anticorrelation between oxygen and sodium abundances in galactic globular clusters \cite{Carretta09-AA} requires for its interpretation a precise knowledge of the production and destruction reactions of oxygen and of the only stable sodium isotope, $^{23}$Na. It is believed that these sites bear the nucleosynthetic imprint of previous generations of stars. This may involve the so-called hot bottom burning in stars on the asymptotic giant branch (AGB) of the Hertzsprung-Russell diagram \cite{Izzard07-AA}, as well as core hydrogen burning in fast rotating massive stars during the main sequence \cite{Decressin07-AA}. 

\begin{figure}[]
\includegraphics[width=\columnwidth]{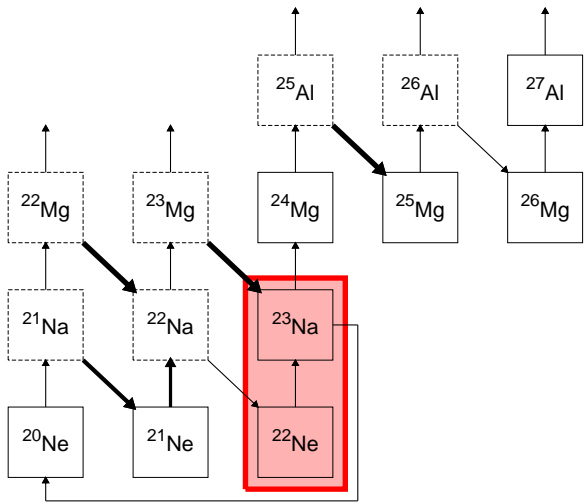}
\caption{\label{fig:NuclideChart} Chart of the nuclides for nuclear reactions on neon and sodium at 80 MK temperature. Stable nuclides are given by solid line boxes, radioactive nuclides by dashed line boxes. The widths of the arrows connecting the nuclides are proportional to the logarithm of the probability for the relevant nuclear transformation (proton capture, radioactive decay, or (p,$\alpha$) reaction), including the recently revised $^{22}$Na(p,$\gamma$)$^{23}$Mg rate \cite{Sallaska10-PRL,Sallaska11-PRC}. 
}
\end{figure}

In these stars, $^{23}$Na may be produced in a hydrogen burning region via the $^{22}$Ne(p,$\gamma$)$^{23}$Na reaction, which is included in the neon-sodium cycle (fig.\,\ref{fig:NuclideChart}). The impact of $^{22}$Ne(p,$\gamma$)$^{23}$Na reaction rate uncertainties on the predicted $^{23}$Na abundance during hot bottom burning has motivated a call for new experimental efforts on the $^{22}$Ne(p,$\gamma$)$^{23}$Na cross section \cite{Izzard07-AA}. Moreover, it was found that models for metal-poor and very metal-poor AGB stars would better match the observations if more $^{23}$Na were to be produced by a higher $^{22}$Ne(p,$\gamma$)$^{23}$Na reaction rate \cite{Doherty14-MNRAS3}. 

A second astrophysical site where the $^{22}$Ne(p,$\gamma$)$^{23}$Na reaction is active are novae \cite{Lundmark21-PASP,Jose98-ApJ}. Nova simulations show a significant impact of the $^{22}$Ne(p,$\gamma$)$^{23}$Na reaction rate on the final $^{22}$Ne abundance \cite{Iliadis02-ApJSS} and possibly on the $^{22}$Ne/$^{20}$Ne isotopic ratio \cite{Jose04-ApJ}. For carbon-oxygen novae models, in addition an impact on the final $^{23}$Na and $^{24}$Mg yields was found \cite{Iliadis02-ApJSS}.

The $^{22}$Ne(p,$\gamma$)$^{23}$Na reaction rate is dominated by a large number of resonances \cite{Rolfs88-Book}. The best known resonance strengths are the ones at $E_{\rm p}$ = 479 and 1279\,keV\footnote{In the present work, $E_{\rm p}$ denotes the proton beam energy in the laboratory system and $E$ the center-of-mass energy.} which are known with 10\% precision each \cite{Longland10-PRC,Keinonen77-PRC}. 
In addition, several low-energy resonances have been studied previously using a windowless gas target with enriched $^{22}$Ne gas, but only experimental upper limits were established \cite{Goerres82-NPA}. No direct resonance strength data exist below $E_{\rm p}$ = 436\,keV. 

The indirect data \cite{Powers71-PRC,Hale01-PRC} and experimental upper limits \cite{Goerres82-NPA} have been treated in different ways in the literature, leading to up to a factor of 1000 revision from the reaction rate reported in the NACRE reaction rate compilation \cite{NACRE99-NPA} to the more recent one by Iliadis {\it et al.} \cite{Iliadis10-NPA841_251}. Still, significant uncertainties of up to a factor of 2.5 remain in the recently recommended reaction rate \cite{Iliadis10-NPA841_251}, clearly calling for new experimental efforts. 

In the present work, a new setup is developed to directly study low-energy resonances in the $^{22}$Ne(p,$\gamma$)$^{23}$Na reaction. 
The resonances addressed in this new setup correspond to a temperature range of approximately 30-500\,MK, thus covering hot bottom burning \cite{Izzard07-AA,Doherty14-MNRAS3}, massive stars \cite{Decressin07-AA}, and novae \cite{Jose98-ApJ,Iliadis02-ApJSS,Jose04-ApJ}.
The setup is located at LUNA (Laboratory for Underground Nuclear Astrophysics) in the Gran Sasso National Laboratory, Italy. LUNA operates the world's only underground accelerator, the 400\,kV LUNA2 machine which has been running uniquely sensitive experiments for a number of nuclear reactions of astrophysical relevance \cite{Costantini09-RPP,Broggini10-ARNPS,Scott12-PRL,Anders14-PRL}. The 1400\,m thick rock overburden above the Gran Sasso underground facility suppresses cosmic muons by six orders of magnitude, leading to unprecedented low $\gamma$-ray background both at low \cite{Caciolli09-EPJA} and at high \cite{Bemmerer05-EPJA,Szucs10-EPJA} $\gamma$-ray energies.

The paper is organized as follows: In section \ref{sec:Feasibility}, a feasibility study using the setup of the previous campaign on the $^2$H($\alpha$,$\gamma$)$^6$Li reaction is described. The $\gamma$-ray background induced by the ion beam is studied. As a concrete verification of the power of underground in-beam $\gamma$-ray spectrometry, the first observation of the $E_{\rm p}$ = 186\,keV resonance using the feasibility study setup is reported. In section \ref{sec:FinalSetup}, the dedicated setup adopted for the study of the $^{22}$Ne(p,$\gamma$)$^{23}$Na reaction is described. Section \ref{sec:Density} describes the measurement of the target density without and with incident ion beam. For the latter purpose, the beam heating effect of protons in neon gas is studied by the resonance scan technique. A summary and an outlook are offered in section \ref{sec:Summary}.

\section{Feasibility study}
\label{sec:Feasibility}

As a first step, a feasibility study of the planned \linebreak $^{22}$Ne(p,$\gamma$)$^{23}$Na experiment was carried out, using the setup from a previous experiment (sec.\,\ref{subsec:d+alpha-Setup}) without any changes.  First, possible nuclear reactions of ion beam background are reviewed (sec.\,\ref{subsec:GammaBackground}). Second, the resonance at $E_{\rm p}$ = 186\,keV is observed for the first time (sec.\,\ref{subsec:Reso186}). The feasibility study informed the development of a new dedicated setup (sec.\,\ref{sec:FinalSetup}).

\subsection{Experimental setup for the feasibility study}
\label{subsec:d+alpha-Setup}

In order to study the origin and intensity of possible ion beam induced background, the setup from the previous LUNA experimental campaign on the $^2$H($\alpha$,$\gamma$)$^6$Li reaction  \cite{Anders13-EPJA,Anders14-PRL} was used. The target chamber was a box-shaped steel container of 11 $\times$ 13\,cm inner area, 44\,cm length and 2\,mm thickness and an end flange for the calorimeter. Six cm downstream of the calorimeter flange, a 13\,cm wide and 4.4\,cm deep recess was included in the target chamber for the high-purity germanium (HPGe) detector \cite{Anders13-EPJA}. The HPGe detector used had 137\% relative efficiency\footnote{The relative efficiency is obtained by comparing the counting rate of the 1.33\,MeV $^{60}$Co line at 25\,cm distance from the end cap with the same rate in a 3''$\times$3'' cylindrical sodium iodide detector.} and was positioned with its end cap just 1.5\,cm below the beam axis.

\begin{figure*}[]
\includegraphics[width=0.5\textwidth]{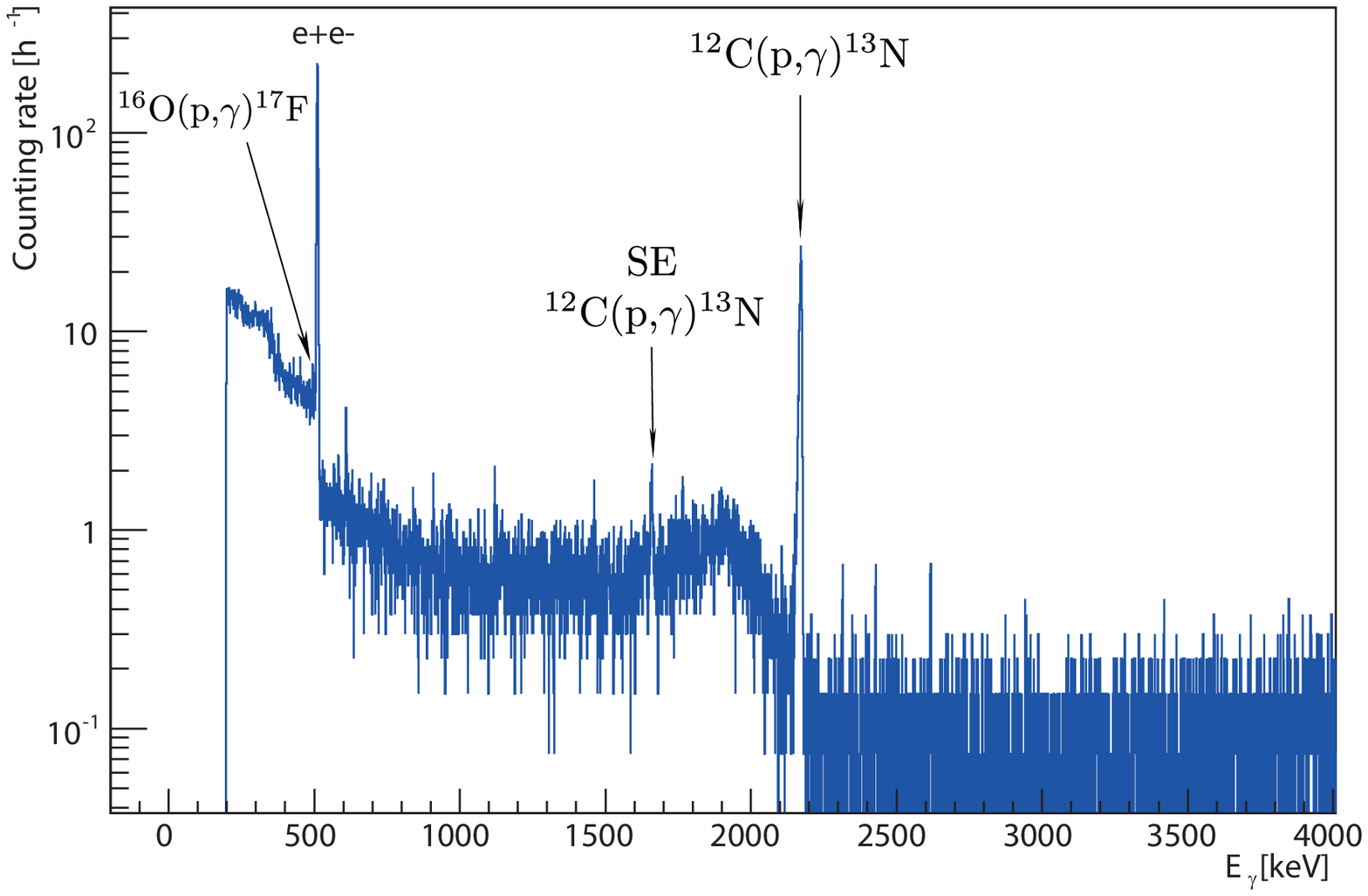}%
\includegraphics[width=0.5\textwidth]{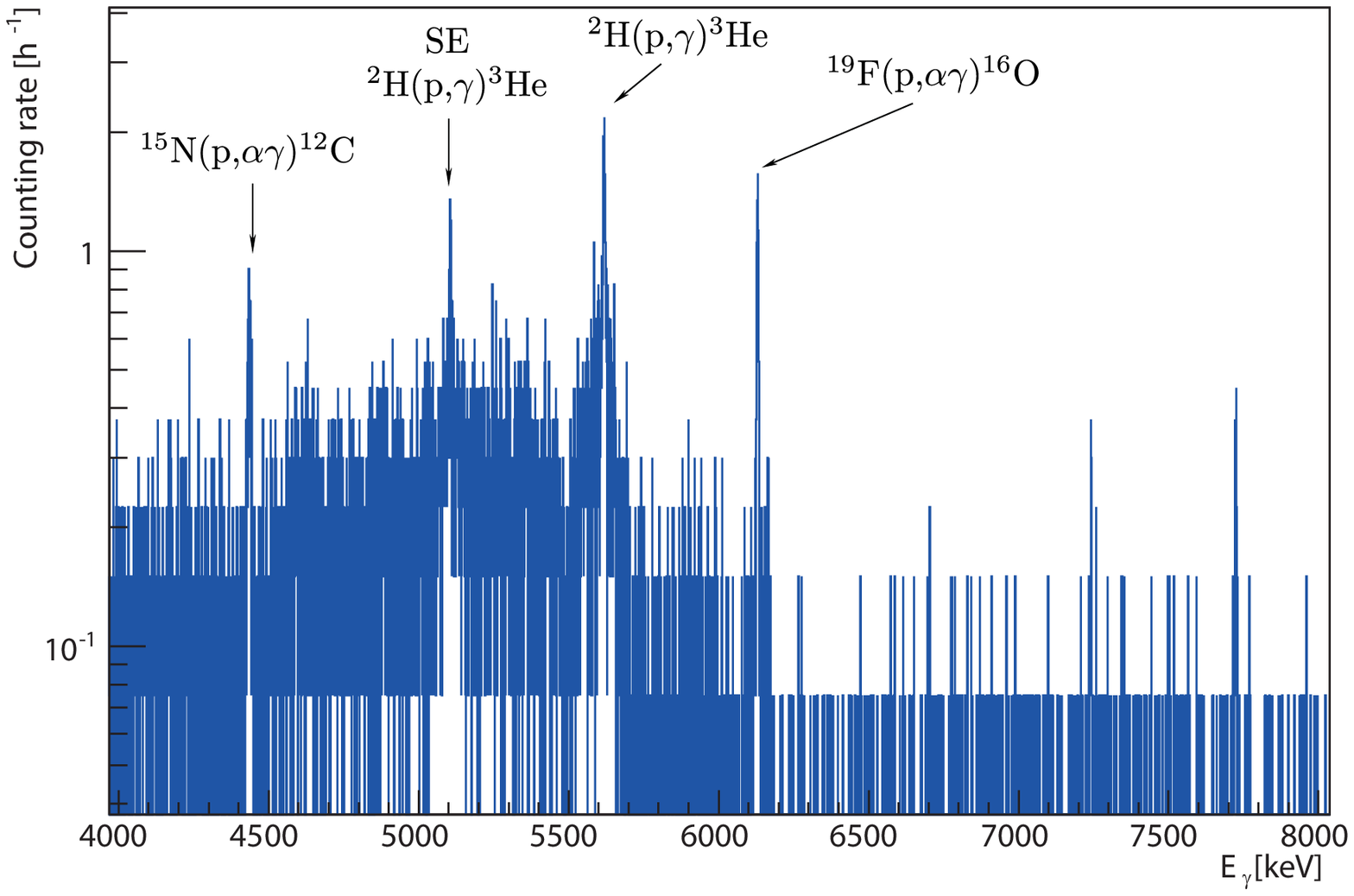}
\caption{\label{fig:GePDSpectrum} Observed $\gamma$-ray spectrum with the HPGe detector, at a beam energy $E_{\rm p}$ = 258\,keV.}
\end{figure*}

In the center of the target chamber there was a rectangular steel inner tube (1.8$\times$1.8$\times$17\,cm$^3$, 0.1\,cm wall thickness). In the $^2$H($\alpha$,$\gamma$)$^6$Li campaign, its purpose was to limit the interaction of elastically scattered target deuterons with the rest of the target gas, reducing the production of neutrons from the $^2$H(d,n)$^3$He reaction. This tube had no effect on the present study but was kept in order to conserve the integrity of the setup.

\subsection{$\gamma$-ray background induced by the ion beam}
\label{subsec:GammaBackground}

The background induced by cosmic rays at Gran Sasso National Laboratory is negligible due to the thick rock overburden \cite{Broggini10-ARNPS}. In order to also shield against environmental radionuclides, the setup was surrounded by a 20\,cm thick lead shield  \cite{Anders13-EPJA}. Only a small counting rate from $\gamma$ lines originating from $^{40}$K and from the $^{238}$U and $^{232}$Th decay chains remains (fig.\,\ref{fig:GePDSpectrum} and Ref.\,\cite{Anders13-EPJA}).

For the study of the ion beam induced background, neon gas with natural isotopic composition (9.25\% $^{22}$Ne) was used in recirculation mode. The proton beam energy range explored was $E_{\rm p}$ = 120-400\,keV, with target pressures of $p_{\rm T}$ = 0.6-2.5\,mbar. As the target chamber was previously used with deuterium gas during the $^2$H($\alpha$,$\gamma$)$^6$Li campaign it was designed for, the data are expected to give a pessimistic worst-case estimate of what ion beam induced background to expect.

As an example, a test run at $E_{\rm p}$ = 258\,keV is shown in fig.\,\ref{fig:GePDSpectrum}. This energy is below the $E_{\rm p}$ = 271.6 keV \linebreak $^{21}$Ne(p,$\gamma$)$^{22}$Na resonance which would otherwise dominate the spectrum (see fig.\,\ref{fig:NaISpectrum}, lower spectrum) but that will not be present in isotopically enriched $^{22}$Ne gas. However, it is still strong enough to give a reasonable yield for other parasitic lines. At a proton current of $I$ = 126\,$\mu$A, a total accumulated charge of 6\,C and 0.8\,mbar Ne target gas, the monitor run showed a number of ion beam induced lines above background (fig.\,\ref{fig:GePDSpectrum}). The lines and their likely sources are listed in Table\,\ref{table:Reactions}, ordered by the $\gamma$-ray energy.

Due to implantation of deuterium gas in surfaces of the chamber and calorimeter during the long running times of the previous $^2$H($\alpha$,$\gamma$)$^6$Li experiment \cite{Anders13-EPJA}, there is an ample deuterium target for the $^{2}$H(p,$\gamma$)$^{3}$He reaction ($Q$-value $Q$ = 5493\,keV). This reaction leads to a single $\gamma$ ray at energy $Q+2/3 \times E_{\rm p}$. The target chamber and collimator will be replaced for the future experiment, so this reaction is not expected to present a challenge.

The $^{12}$C(p,$\gamma$)$^{13}$N reaction ($Q$ = 1943\,keV) originates from hydrocarbons present in the vacuum and then adsorbed onto metallic surfaces. It gives rise to a single $\gamma$ ray at energy $Q+12/13 \times E_{\rm p}$ that is well visible in the low-energy part of the spectrum. This contaminant is expected to be initially lower in the planned new setup and then to slowly grow with time. The same is true for the $^{13}$C(p,$\gamma$)$^{14}$N reaction ($Q$ = 7551\,keV, isotopic enrichment of $^{13}$C 1.07\%), which at LUNA energies also gives rise to a single $\gamma$ ray at energy $Q+13/14 \times E_{\rm p}$.

There are two reactions that may in principle populate the first excited state of $^{12}$C at 4439\,keV: the $^{15}$N(p,$\alpha\gamma$)$^{12}$C  ($Q$ = 4965\,keV) and $^{11}$B(p,$\gamma$)$^{12}$C  ($Q$ = 15957 keV) reactions. The latter leads to a $\gamma$ ray at 11.7\,MeV but usually plays a role only near its $E_{\rm p}$ = 163\,keV resonance. Therefore it is suspected here that some implanted nitrogen impurities, incurred during test phases with nitrogen gas, are the origin for these $\gamma$ rays. 

The $^{16}$O(p,$\gamma$)$^{17}$F reaction ($Q$ = 600\,keV) gives rise to the very weak $\gamma$ ray at 495\,keV from the decay of the first excited state of $^{17}$F. It is believed that the oxygen is in the oxidized surface of the copper head of the calorimeter. Due to its very low impact, no measures are needed to mitigate this background.

The well-known $^{19}$F(p,$\alpha\gamma$)$^{16}$O background reaction \linebreak gives rise to a $\gamma$ ray at 6130\,keV with a particular structure due to stopped and Doppler-shifted $\gamma$ rays from the decay of the 6130\,keV excited state in $^{16}$O. The fluorine is an ingredient of the heat conducting paste used in the calorimeter end cap, and it is supposed that some additional fluorine originates from the Viton O-rings placed at several locations in the windowless gas target setup. The particular beam stop used for the test experiment had been in contact with heat conducting paste several times. For the future experiment, a beam stop that is made from one piece and does not need heat conducting paste will be used, hoping to reduce the fluorine contamination.

In conclusion, the main problem identified here is the $^{19}$F(p,$\alpha\gamma$)$^{16}$O reaction. It will also be relevant to the experiment with enriched gas. From yield measurements and excitation functions with and without gas, it was found that the $^{19}$F contaminant is mainly located on the calorimeter end cap, and to a lower level on the target entrance collimator (named AP$_1$ in sec.\,\ref{sec:FinalSetup}). These are the critical surfaces, where fluorine will be removed by mechanical and chemical cleaning in order to reduce that particular background. 
 
\begin{table}[bt]
\resizebox{0.5\textwidth}{!}{%
\begin{tabular}{|r|l|p{5cm}|} \hline
\bf $E_\gamma$ [keV] & Nuclear reaction & Place of origin\\ \hline
495  & $^{16}$O(p,$\gamma$)$^{17}$F & Oxidized apertures and calorimeter end cap \\
2168 & $^{12}$C(p,$\gamma$)$^{13}$N & Calorimeter end cap \\
4439  & $^{11}$B(p,$\gamma$)$^{12}$C & Target \\
4439  & $^{15}$N(p,$\alpha\gamma$)$^{12}$C & Implanted nitrogen \\
5617 & $^{2}$H(p,$\gamma$)$^{3}$He & Calorimeter end cap \\
6130 & $^{19}$F(p,$\alpha\gamma$)$^{16}$O & Calorimeter end cap\\
     & & (heat conducting paste) \\
 \hline
\end{tabular}
}
\caption{\label{table:Reactions} List of $\gamma$ rays induced by the ion beam at $E_{\rm p}$ = 258\,keV (fig.\,\ref{fig:GePDSpectrum}). The relevant nuclear reaction and its probable place of origin are given.
}
\end{table}

\subsection{First observation of the 186\,keV resonance}
\label{subsec:Reso186}

\begin{figure}[b]
\includegraphics[width=\columnwidth]{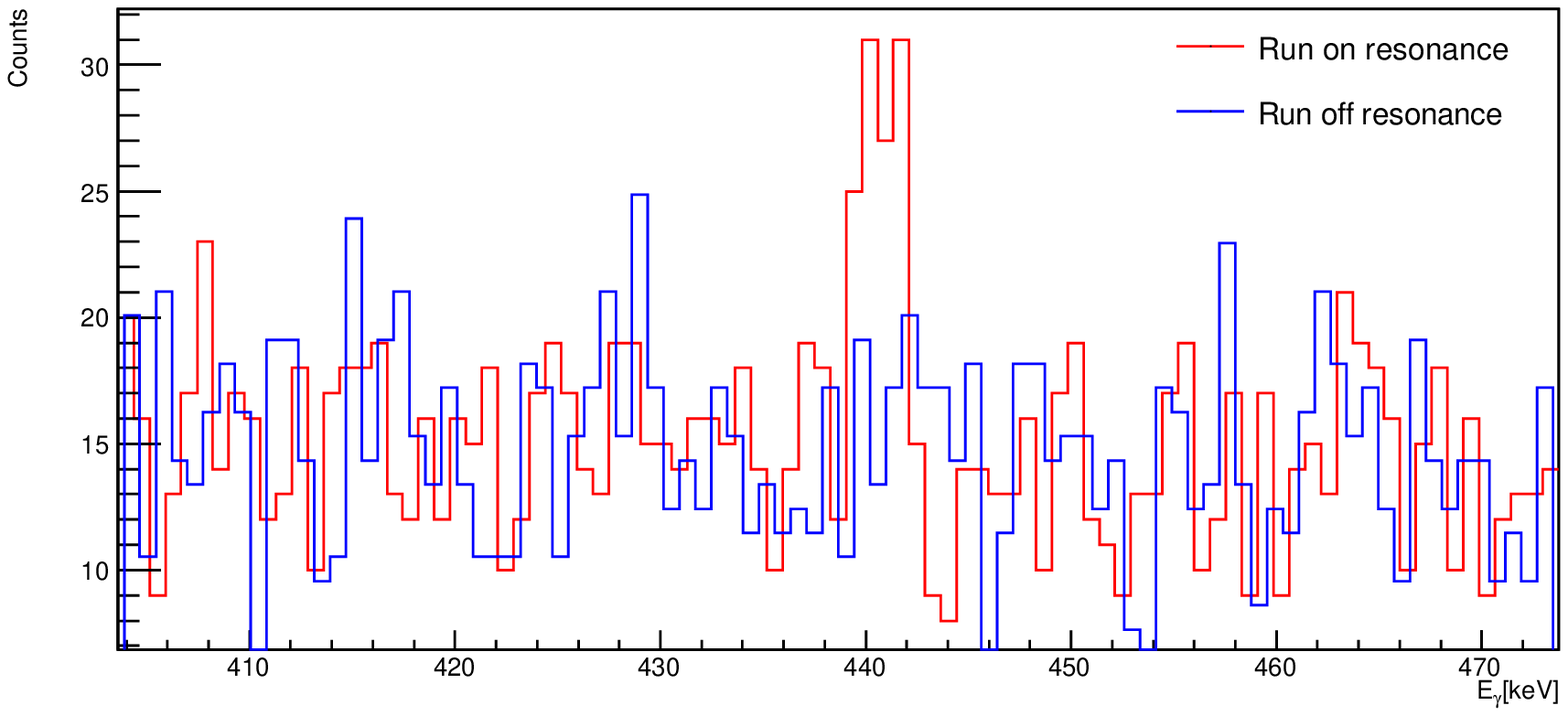}
\includegraphics[width=\columnwidth]{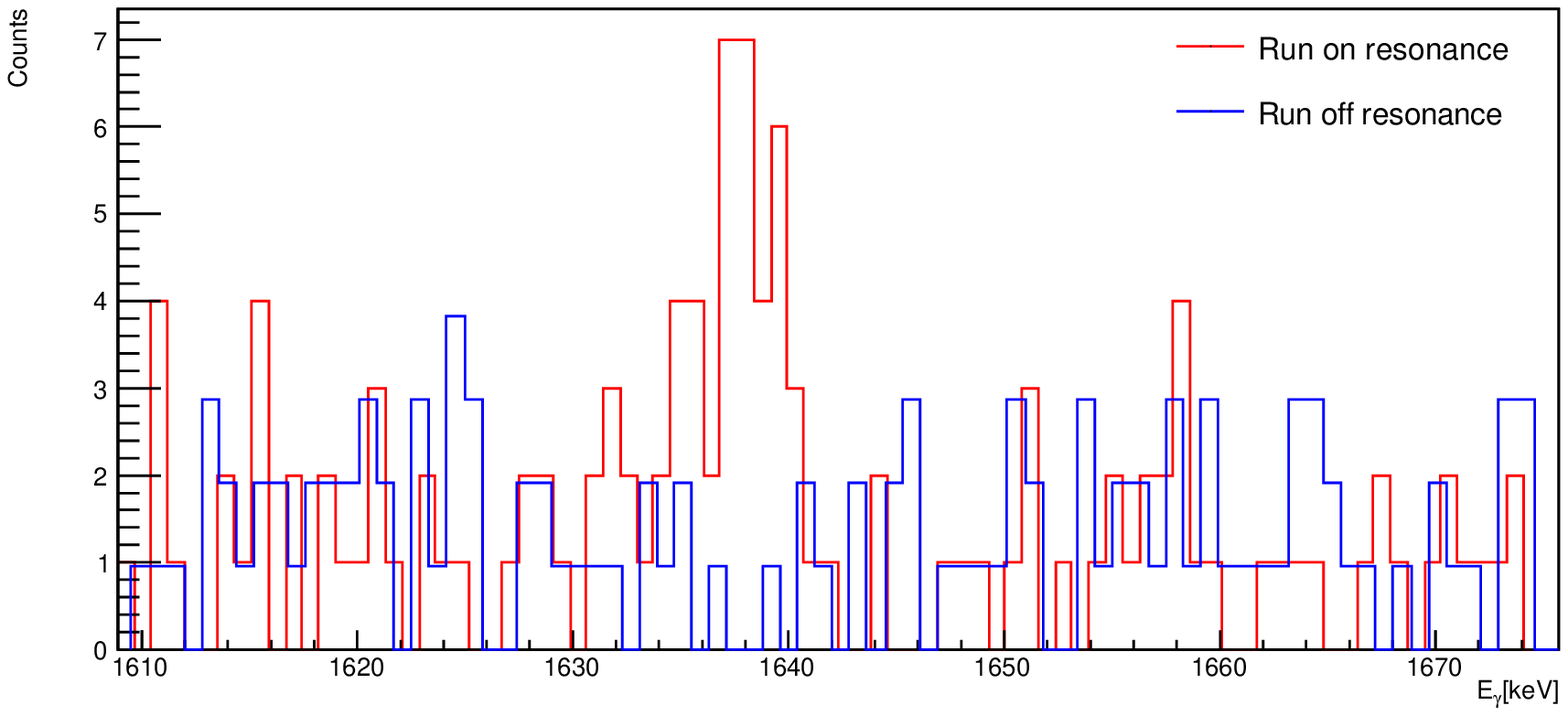}
\caption{\label{fig:GePDSpectrum_186} Observed $\gamma$-ray spectra near 440 (top panel) and 1636\,keV (bottom panel), for the $E_{\rm p}$ = 186\,keV resonance in $^{22}$Ne(p,$\gamma$)$^{23}$Na. The off-resonance spectrum is overplotted, scaled by a factor of 0.96 for the same charge.}
\end{figure}

In order to further check the feasibility of the chosen approach by studying a concrete example, in an overnight run the hypothetical $^{22}$Ne(p,$\gamma$)$^{23}$Na resonance at $E_{\rm p}$ = (186$\pm$3)\,keV was studied. This resonance corresponds to the $E_{\rm x}$ = 
(8972$\pm$3)\,keV excited state in $^{23}$Na. For the strength of this resonance, only an experimental upper limit is available in the literature \cite{Goerres82-NPA} and adopted in the reaction rate compilations \cite{NACRE99-NPA,Iliadis10-NPA841_251}.

Also this run was performed with the setup of the previous $^2$H($\alpha$,$\gamma$)$^6$Li experiment (sec.\,\ref{subsec:d+alpha-Setup}). A proton beam of $E_{\rm p}$ = 191\,keV with 115\,$\mu$A beam intensity was sent onto the gas target with $p_{\rm T}$ = 1.5\,mbar natural neon gas inside. A total charge of 5.3\,C was collected. 

The $\gamma$ rays at 1636\,keV (from the decay of the second excited state to the first excited state in $^{23}$Na) and at 440\,keV (decay of the first excited state in $^{23}$Na) have both been observed (fig.\,\ref{fig:GePDSpectrum_186}). The net number of counts is 
43$\pm$17 for 440\,keV and 25$\pm$10 for the 1636\,keV line. A short off-resonance spectrum at $E_{\rm p}$ = 161\,keV has been taken in order to verify whether the observed peaks are due to direct capture. None of the two peaks are found in the off-resonance spectrum, confirming that the peaks are due to the resonance. The off-resonance run has also been repeated with the same charge as the on-resonance run but at 1.0\,mbar target pressure, and again no signal was observed at 440 and 1636\,keV (fig.\,\ref{fig:GePDSpectrum_186}).

For the simple demonstration of the existence of the resonance, instead of a full scan of the resonance profile just an on-resonance and an off-resonance run with similar statistical significance are needed. However, it is not possible to know whether the resonance was completely populated and at which point in the setup the point of maximum emission is located. Both of these aspects can potentially lead to an underestimation of the resonance strength. 

Assuming a branching ratio of 100\% for 440\,keV and 92\% for the 1636\,keV  $\gamma$ ray \cite{Firestone07-NDS23}, and using the maximum of the detection efficiency \cite{Anders13-EPJA}, i.e. population of the resonance directly at the center of the target, a lower limit for the resonance strength of
\begin{equation}
\omega \gamma (E_{\rm p}=186\,{\rm keV}) \geq 0.12 \times 10^{-6}\,{\rm eV\, (90\,\%\,C.L.)}
\end{equation}
is obtained. This lower limit is to be compared to the previous experimental upper limit of 2.6\,$\times$\,10$^{-6}$\,eV \cite{Goerres82-NPA}. 

The corresponding level has previously been observed in the $^{22}$Ne($^3$He,d)$^{23}$Na reaction \cite{Powers71-PRC,Hale01-PRC}. In neither of the two works, a clear $J^{\Pi}$ assignment was possible based on the observed angular distributions. Assuming $J^{\Pi}$ = 5/2$^+$, a resonance strength of 3.4\,$\times$\,10$^{-6}$\,eV was found \cite{Hale01-PRC}, but the same authors then opted to adopt the previously mentioned upper limit \cite{Goerres82-NPA} instead. 

Due to its energy, this resonance has the largest impact on the reaction rate near 180\,MK. This temperature is important for hot bottom burning \cite{Izzard07-AA} and novae \cite{Jose98-ApJ}. The lower limit developed here corresponds to a reaction rate that is 6\% of the total thermonuclear reaction rate in the Iliadis {\it et al.} compilation at this temperature. If one were to assume instead 1.4\,$\times$\,10$^{-6}$\,eV, the average of the present lower limit and the previous upper limit \cite{Goerres82-NPA}, the present resonance would contribute 60\% of the total Iliadis {\it et al.} rate at this temperature.

The final value for the resonance strength will be determined in the future using the dedicated setup discussed below and enriched $^{22}$Ne gas.

The upper limits from the literature \cite{Goerres82-NPA} for several other resonances in the LUNA energy range are similar to the one of the 186\,keV resonance, so it is likely that also in these cases either a greatly improved upper limit or a positive detection can be achieved by this basic approach. These resonances directly impact the overall reaction rate in the LUNA energy range \cite{Hale01-PRC}.

The feasibility study so far, including especially the first observation of a new resonance, has already shown that the basic approach of an in-beam experiment with a HPGe detector and a windowless gas target is sound. Even still, some adjustments will improve the prospects for the dedicated study of $^{22}$Ne(p,$\gamma$)$^{23}$Na. These are discussed in the following section.

\section{New, dedicated experimental setup for the $^{22}$Ne(p,$\gamma$)$^{23}$Na experiment}
\label{sec:FinalSetup}

Three modifications in particular are required for further improvements. 
First, for the enriched noble gas $^{22}$Ne the gas consumption must be reduced by using a recirculating windowless gas target system. Its salient features are described below (sec.\,\ref{subsec:DiffPumpedGasTargetSystem}). Second, due to a number of unobserved resonances with sometimes unknown spin-parity, the possible effects of $\gamma$-ray angular distributions must be mitigated by adding a second HPGe detector at a different angle. This is achieved by placing a HPGe detector at 90$^\circ$ in downlooking geometry (sec.\,\ref{sec:TwoHPGeSetup}). Third, the germanium detectors must be well collimated so that the experimental yield is dominated by the narrow resonances to be studied, not by the continuum outside the resonances. 

In the following, the new improved setup and preliminary measurements to characterize the gas target  are described. 

\subsection{Differentially pumped gas target system and beam calorimeter}
\label{subsec:DiffPumpedGasTargetSystem}

\begin{figure*}[]
\includegraphics[width=\textwidth]{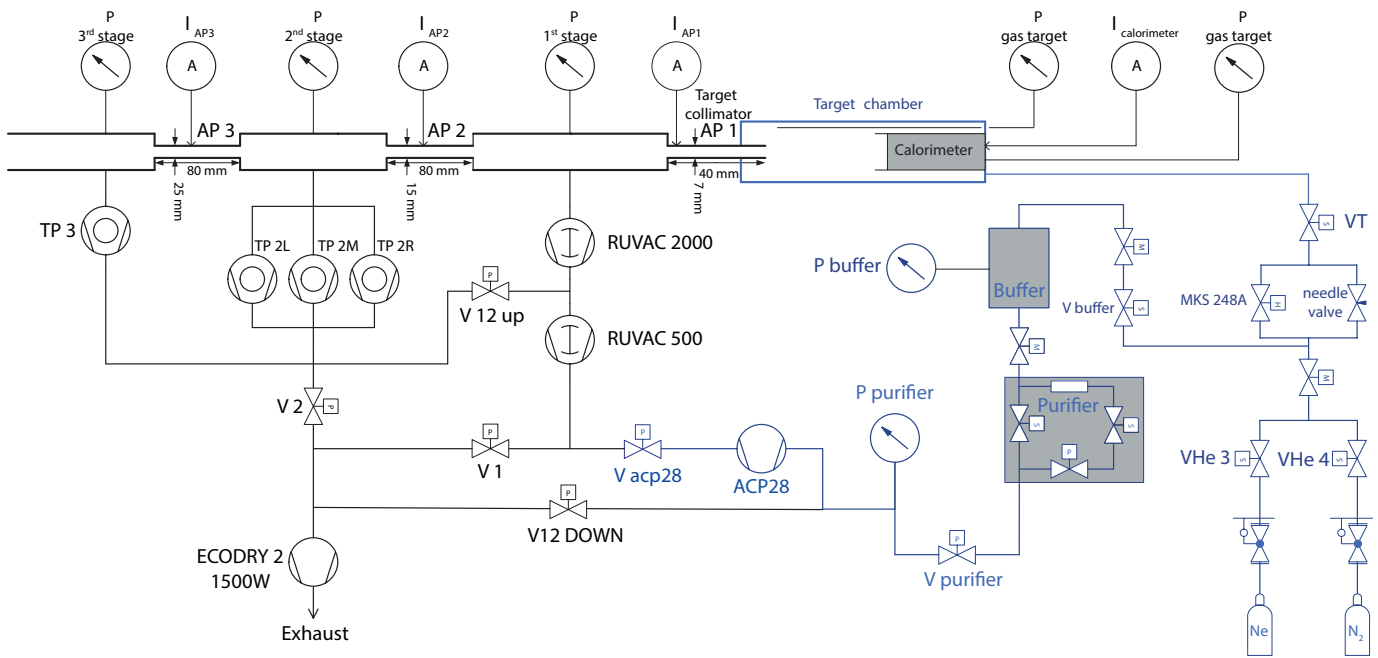}
\caption{\label{fig:Pumps} Scheme of the differential pumping system used for the experiments. In recirculating mode, the valves V$_1$, V$_2$, and V$_{\rm 12\,down}$ are closed. The changes with respect to the previous version \cite{Anders13-EPJA} are shown in blue.}
\end{figure*}

The windowless, differentially pumped gas target system consists of three pumping stages (fig.\,\ref{fig:Pumps}). The gas is inserted into the target chamber through valve V$_{\rm T}$ and then flows into the direction of the accelerator tank (from the right to the left in fig.\,\ref{fig:Pumps}) through the water-cooled target collimator (hereafter called AP$_1$; diameter $d_1$ = 7\,mm, length $l_1$ =  40\,mm). This aperture is designed to be sufficiently long and narrow to enable a typical pressure drop of a factor of 100 between target chamber and first pumping stage (sec.\,\ref{subsec:PressTempProfile}). 

The first pumping stage is equipped with a 2050\,m$^3$/h Roots pump, backed by a 500\,m$^3$/h  Roots pump, which maintains a medium vacuum of typically 10$^{-3}$\,mbar. Typically more than 99\% of the gas flow from the target chamber is carried away by the Roots pumps of the first pumping stage. The remaining gas may then pass aperture AP$_2$ ($d$ = 15\,mm, $l$ = 80\,mm) and enters the second pumping stage, equipped with two 1000\,l/s and one 1500\,l/s turbomolecular pumps. The small quantity of gas still remaining  may then pass aperture AP$_3$ ($d$ = 25\,mm, $l$ = 80\,mm) and enter the third pumping stage, which is evacuated by a 360\,l/s turbomolecular pump. In routine operations, also the turbomolecular pumps of the second and third pumping stages are backed by the 500\,m$^3$/h Roots pump through valve V$_{\rm 12\,up}$. Typical pressures in the second and third pumping stages are in the 10$^{-6}$ - 10$^{-7}$\,mbar range. 

When isotopically enriched or rare gases are used, the exhaust from the 500\,m$^3$/h Roots pump cannot be discarded but must instead be used again for the experiment. In that case, valves V$_1$, V$_2$, and V$_{\rm 12\,down}$ are closed (fig.\,\ref{fig:Pumps}), and the gas target system is used in recirculation mode. Then, the exhaust from the 500\,m$^3$/h Roots pump is compressed by a dry 27\,m$^3$/h forepump (ACP28) and sent to a chemical getter removing oxygen and nitrogen contaminations (Monotorr II purifier). Downstream of the purifier, a buffer volume of 1\,liter with a pressure of 200-800\,mbar is kept and the gas is then re-inserted into the target chamber.

The pressure in the target chamber is measured independently by two capacitance manometers (precision 0.25\%) that are connected by tubes to the front and the back end of the target chamber, respectively. The back end manometer (MKS baratron 626A) is also used to control, via an analog feedback unit, part of the gas flow reaching the target chamber via a thermal leak valve (MKS 248A). In order to better exploit the dynamic range of the MKS 248A, a constant offset gas flow is added via a manually controlled needle valve. The value of the target pressure is logged every second; the other pressure values are logged every 5 seconds. The target pressure was found to be constant to better than 1\% in normal operations. 

A separate inlet to the target chamber filling system allows fresh neon gas to be filled in from a pressurized container. Alternatively, nitrogen gas can be used, in order to avoid introducing water vapor when shutting down the pumps and restoring the system to atmospheric pressure.

All the pumps and above mentioned valves except for the needle valve and the MKS 248A are controlled via a Labview Fieldpoint 2000 system that is, in turn, remotely controlled via a personal computer in the control room. The Labview system reads and logs all relevant pressure values. 

The ion beam arrives in the gas target system from the left side of fig.\,\ref{fig:Pumps}. It passes apertures AP$_3$, AP$_2$, and AP$_1$ and then enters the gas target chamber. The electrical currents on all three apertures are read out via analog ammeters located in the control room. When there is no gas in the target, the aperture current readings are used to optimize the ion beam transmission to the target. When there is gas inside the target, the electrical readings become meaningless due to the large contribution of secondary electrons. 

The ion beam is stopped on the copper head of a beam calorimeter, which maintains a constant temperature gradient between a hot and a cold side at temperatures 70\,$^\circ$C and 0\,$^\circ$C, respectively. The cold side is cooled by a refrigerating system with a thermostat. The hot side is heated by thermoresistors. The temperatures at several points of the calorimeter are read out via a second Labview Fieldpoint 2000 system (i.e. in addition to the one controlling the gas target pumps, valves, and pressures). These values are then used in a software-controlled feedback system to control the heating power, maintaining the hot side at constant temperature. The constant temperature gradient ensures that the heating power $W$ needed to maintain the gradient is also constant. The calorimeter design is based on a previous version \cite[for more details]{Casella02-NIMA} and has been optimized for beam powers in the 10-140\,W range.

The so-called zero power without ion beam $W_0$ is determined by measuring the power on the heating resistors over a period of several minutes with the gas in the target. Then, the beam intensity $I$ (in units of electrical current of singly charged H$^+$ ion beam) is given by 
\begin{equation} \label{eq:Calorimeter}
I = C_{\rm calo} \frac{W_{0}-W_{\rm run}}{E_{\rm p} - \Delta E_{\rm p}^{\rm target}}
\end{equation} 
where $W_{\rm run}$ is the heating resistor power measured with ion beam on the target, $E_{\rm p}$ is the projectile energy in the laboratory system calculated from the accelerator energy calibration \cite{Formicola03-NIMA}, and $\Delta E_{\rm p}^{\rm target}$ is the energy loss of the projectile over the length of the gas target. $C_{\rm calo}$ $\approx$ 1 is the calorimeter calibration constant and has to be determined empirically.  The energy loss $\Delta E_{\rm p}^{\rm target}$ depends on the well-known stopping power tables by SRIM \cite{Ziegler10-NIMB}, and on the target density profile (secs.\,\ref{subsec:PressTempProfile} and \ref{subsec:BeamHeating}). In cases where the beam heating effect significantly affects the target density (and thus again the beam heating effect), eq. (\ref{eq:Calorimeter}) is used iteratively (one iteration proved sufficient) to remove the degeneracy between $I$ and $\Delta E_{\rm p}^{\rm target}$.

\subsection{Target chamber with two germanium detectors}
\label{sec:TwoHPGeSetup}

\begin{figure}[b]
\includegraphics[width=\columnwidth, angle=0]{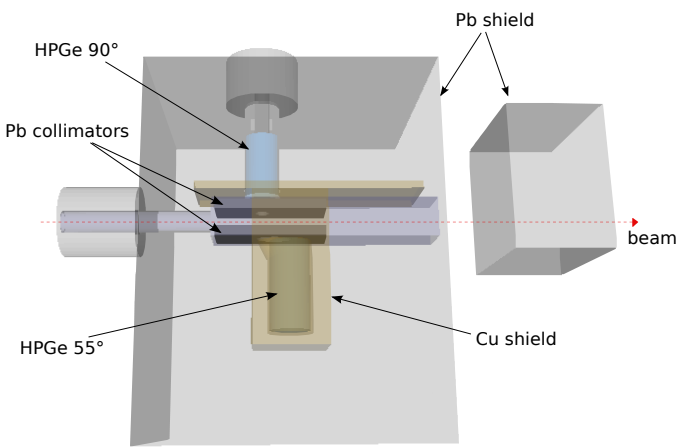} 
\caption{\label{fig:SimulatedSetup} Schematic view of the experimental setup adopted for the planned $^{22}$Ne(p,$\gamma$)$^{23}$Na experiment.}
\end{figure}

The dedicated setup for the study of the $^{22}$Ne(p,$\gamma$)$^{23}$Na reaction includes a windowless gas target chamber and two HPGe detectors (fig.\,\ref{fig:SimulatedSetup}).  

The target, pumping stages, and entry collimator have already been described in sec.\,\ref{subsec:DiffPumpedGasTargetSystem}. The gas target will be used with $^{22}$Ne gas (isotopic enrichment 99.99$\%$) in recirculation mode, with a chemical getter removing impurities that may diffuse into the target area. Between the end of the 4\,cm long entry collimator (diameter 7\,mm), on the upstream side and the calorimeter end cap, on the downstream side, the gas target chamber is a box of 33.0$\times$12.0$\times$10.4\,cm$^3$ (length $\times$ width $\times$ height) volume. 

Gamma rays are detected by two HPGe detectors, each specially manufactured to satisfy ultra low background (ULB) specifications. The first one, with 137\% relative efficiency, is located below the target chamber, at an effective angle of 55$^\circ$. The second one, with 90\% relative efficiency, is located above the target chamber at an angle of 90$^\circ$. By comparing the counting rates from the two detectors, an upper limit can be placed on possible effects of $\gamma$-ray angular distributions. 

Inside the gas target chamber, $\gamma$-ray collimators are placed: at the bottom of the chamber, a 2.8\,cm thick lead brick with a hole shaped as a truncated cone with elliptical base, the main axis of which is inclined by 45$^\circ$. Behind it, a 1.6\,cm thick tungsten disk is placed to attenuate parasitic $\gamma$ rays originating on the calorimeter end cap. Taken together, these collimators lead to an effective angle of  55$^\circ$  for the lower HPGe detector. 
Near the top of the chamber, a 2.8\,cm thick lead brick with a tapered hole (diameter at the lower end 5.1\,cm and at the upper end 7.0\,cm) is placed on a steel stand, in order to collimate the $\gamma$ rays for the upper, 90$^\circ$ HPGe detector. As the pressure profile is flat inside the target chamber, those inserts are not expected to influence the density profile outside the error bars given below (secs.\,\ref{subsec:PressTempProfile} and \ref{subsec:BeamHeating}).

The two HPGe detectors are shielded against environmental radioactivity by a thick graded shield that was developed on the basis of a shielding that has been described previously \cite[so-called setup C]{Caciolli09-EPJA}, with some modifications to allow for the inclusion of the second HPGe detector at 90$^\circ$. The lower, 55$^\circ$ HPGe detector is surrounded by a 4\,cm thick, inner copper shielding and, subsequently, a 25\,cm thick lead shielding. The upper, 90$^\circ$ HPGe detector will be shielded by just 25\,cm of lead. 

The final layout has been coded in GEANT4 \cite{Agostinelli03-NIMA} 
simulations of the setup, in order to confirm the optimal position for the 90$^{\circ}$ detector and to estimate the environmental background reduction. In the simulations, $^{232}$Th was equally spread outside the setup (on a sphere of 2 meters radius centered on the 55$^\circ$ detector end cap) in order to get a rough understanding of the impact of the environmental background. According to the simulations, the shield will reduce the $\gamma$-ray background near $E_\gamma$ = 440\,keV by a factor of 5000 (200) in the 55$^\circ$ (90$^\circ$) detector. Near $E_\gamma$ = 1636\,keV, the suppression is a factor of 8000 (200) in the 55$^\circ$ (90$^\circ$) detector. 

\section{Study of the effective target density in the new setup}
\label{sec:Density}

After the development of the new experimental setup was concluded (sec.\,\ref{sec:FinalSetup}), a dedicated campaign was undertaken to determine the effective density of the windowless gas target in the relevant regions of the target chamber. For this purpose, two mockup chambers were built (sec.\,\ref{subsec:DensitySetup}). 

Using these experimental setups, the pressure and temperature (thus the density) inside the target were measured without ion beam (sec.\,\ref{subsec:PressTempProfile}). Subsequently, the beam heating effect for protons in neon gas was determined using the resonance scan technique (sec.\,\ref{subsec:BeamHeating}).

\subsection{Setup for studying the gas target density} 
\label{subsec:DensitySetup}

In order to determine the effective target density, the temperature and pressure of the target gas have to be known as a function of the position inside the gas target. The gas target chamber has been designed in such a way that a flat density profile is expected in the region observed by the two HPGe detectors. However, it is still necessary to experimentally verify this expected profile.

\begin{figure}[]
\includegraphics[width=\columnwidth]{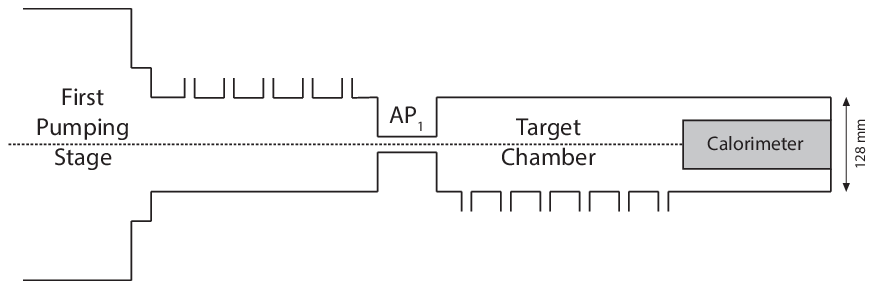}
\includegraphics[width=\columnwidth]{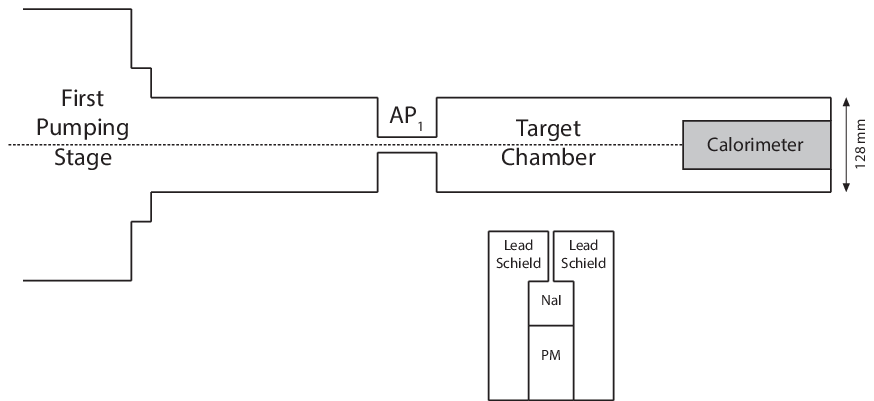}
\caption{\label{fig:SetupFlute}\label{fig:SetupBH} Schematic view of the experimental setups used for the density profile measurement (top) and for the study of the beam heating effect (bottom).}
\end{figure}

A dedicated target chamber and interconnecting pipe have been built with eleven ports with KF16 flanges that can be used to connect pressure and temperature gauges (fig.\,\ref{fig:SetupFlute}, top panel). All other features of this chamber have been designed such that they are equal or very close to those of the final target chamber (sec.\,\ref{sec:FinalSetup}). 

Four different capacitance-based manometers (calibra\-ted to 0.20\% or 0.25\% precision depending on the type) and four Pt100 thermoresistors, mounted on vacuum feed-throughs and long leads in order to place the Pt100 directly along the ion beam axis, have been used for these measurements. 

In order to study the beam heating effect which reduces the effective target density, a similar setup was used (fig.\,\ref{fig:SetupFlute}, bottom). The manometers and thermoresistors were removed, the chamber was replaced with a similar chamber without the connecting flanges, and a collimated 2'' sodium iodide (NaI) detector was placed perpendicular to the beam direction. The detector was surrounded by a 5\,cm thick lead shield, except for a cylindrical hole of 2\,cm diameter at the front face of the shield. The latter hole allows energetic $\gamma$ rays from the $E_{\rm p}$ = 271.6\,keV resonance in the  $^{21}$Ne(p,$\gamma$)$^{22}$Na reaction to enter. The NaI detector and its shielding were placed on a movable table, so as to vary its position along the beam direction \cite{Menzel12-Diplom}.
 
\subsection{Pressure and temperature profiles}
\label{subsec:PressTempProfile}

For the measurement of the density profile without ion beam, aperture AP$_1$ was water cooled to a constant temperature of 290\,K (17\,$^\circ$C), and the hot side of the calorimeter was kept at 343\,K (70\,$^\circ$C).

Using the setup with special flanges for the pressure measurement (sec.\,\ref{subsec:DensitySetup}, fig.\,\ref{fig:SetupFlute}, top panel), the pressure profile in the target chamber was established. It was not possible to simultaneously cover all flanges with the limited number of manometers available. Therefore, different runs were linked by keeping one or two of the measuring devices at a fixed position and using the remaining ones to map the target chamber and the connecting pipe.

\begin{figure}[]
\includegraphics[width=\columnwidth]{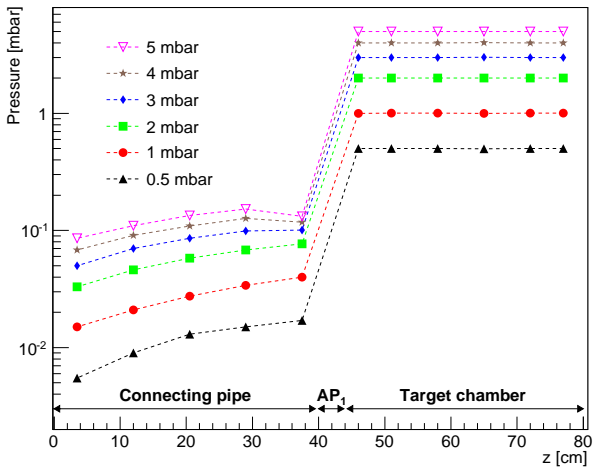}
\caption{\label{fig:PressureProfile} Pressure inside the gas target chamber as a function of the distance $z$ from aperture AP$_2$, for various values of the target pressure $p_{\rm T}$. The lines are just to guide the eye.  }
\end{figure}

The pressure data show a flat profile within the target chamber, constant to better than 1\% (fig.\,\ref{fig:PressureProfile}). The only exception is the point at $z$ = 65\,cm, for which the capacitance manometer used (MKS Baratron 626A) showed a calibration that was (1.8$\pm$0.5)\% lower than the other manometers, consistent over a wide pressure range and several different runs. The pressure for this device was then corrected for this different calibration and higher uncertainty. Except for this one device, all of the data points are mutually consistent, and no difference from run to run was observed for overlapping measurements.

Inside collimator AP$_1$, the pressure dropped by a factor of 30  (fig. \ref{fig:PressureProfile}). Assuming that the pressure drops linearly inside this 4\,cm long aperture, the amount of gas inside it is estimated to be 6\% of the target thickness inside the target chamber itself. 

The temperature profile has been measured in a similar fashion to the pressure profile, using four Pt100 thermoresistors instead of the capacitance manometers. The Pt100 sensors were mounted at the end of long wires ensuring that they were positioned at the center of the gas target chamber. The sensors were calibrated to 0.3\,K, and an additional uncertainty of 1.1\,K due the orientation of the Pt100 sensitive surface (i.e. facing or not the hot side of the calorimeter) was determined by reproducibility tests. 

\begin{figure}[]
\includegraphics[width=\columnwidth]{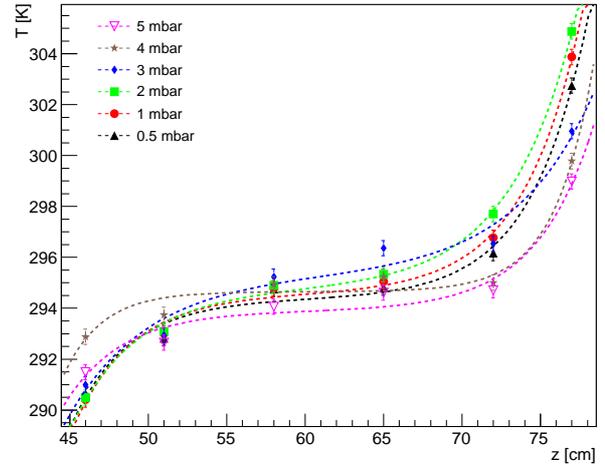}
\caption{\label{fig:TemperatureProfile} Temperature profile inside the gas target chamber as a function of distance $z$ from aperture AP$_2$, for various values of the target pressure $p_{\rm T}$. The curves represent an empirical parameterization. 
}
\end{figure}

The gas temperature was found to increase monotonically from the water cooled AP$_1$ collimator to the hot side of the calorimeter (fig.\,\ref{fig:TemperatureProfile}). When comparing the temperature values at $z$ = 72-77\,cm, near the hot side of the calorimeter, for various pressures, an interesting trend is observed: From 0.5 to 2\,mbar target pressure, the temperature profile gets  steeper with increasing pressure. At 3\,mbar, forced convection \cite{Bird02-Book} starts to contribute significantly, reversing this trend by transporting heat more efficiently in the first few centimeters of gas after the hot side of the calorimeter.

The gas density profile without beam has been determined with an uncertainty of 1.0\% from the pressure and 0.4\% from the temperature, leading to a combined uncertainty of 1.1\%.

\subsection{Study of the beam heating effect}
\label{subsec:BeamHeating}

The density of the target gas may decrease along the beam path because of heat transfer from the intense ion beam. This effect has been studied in detail previously in static gas targets of $^{14}$N gas, using the resonance scan method \cite{Goerres80-NIM,Bemmerer06-NPA}, and of $^3$He gas, using double elastic scattering \cite{Marta06-NIMA}. In the present work, again the resonance scan method is used.

The very narrow ($\Gamma$ $<$ 3\,eV \cite{Becker92-ZPA}) resonance at $E_{\rm p}$ = 271.6\,keV in the $^{21}$Ne(p,$\gamma$)$^{22}$Na reaction is used for the beam heating study. For practical reasons, natural neon gas was used. $^{\rm nat}$Ne contains only 0.27\% of $^{21}$Ne, but the large resonance strength of 82\,meV \cite{Goerres82-NPA} ensures that there is still a satisfactory yield in the $\gamma$-ray detector. 

\begin{figure*}[]
\begin{center}
\includegraphics[height=8cm]{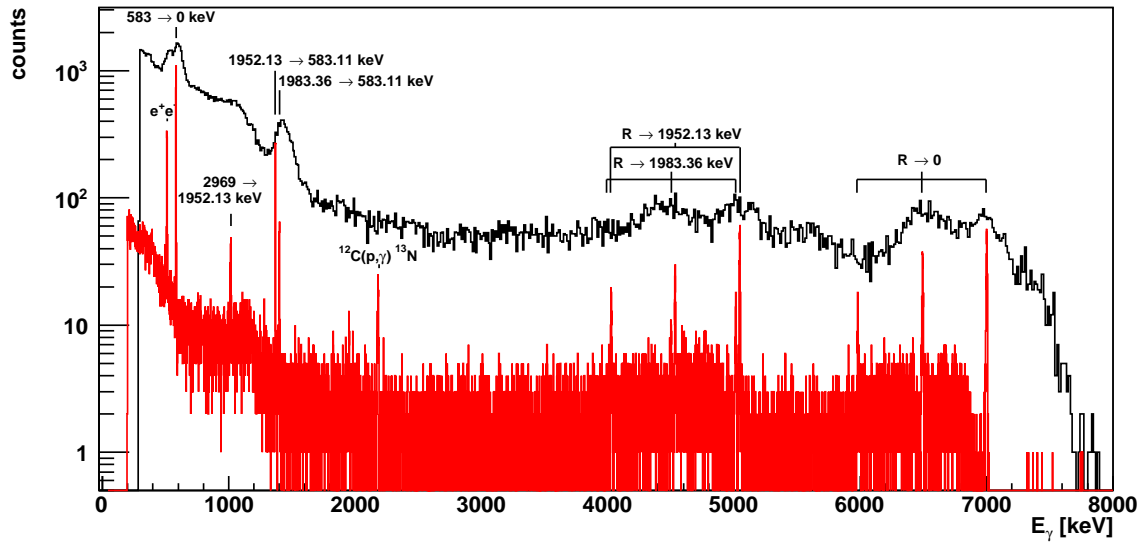}
\caption{\label{fig:NaISpectrum} Upper spectrum: Run with the collimated NaI detector at the $E_{\rm p}$ = 271.6\,keV resonance in the $^{21}$Ne(p,$\gamma$)$^{22}$Na reaction, in a long overnight run. Lower Spectrum: Run with the HPGe detector during the test phase on top of the same resonance.}
\end{center}
\end{figure*}

The $\gamma$-ray spectrum obtained at the $E_{\rm p}$ = 271.6 keV $^{21}$Ne(p,$\gamma$)$^{22}$Na resonance with the NaI detector is dominated by the strong secondary $\gamma$-rays at 583 and 1369\,keV and by a significant continuum up to 8\,MeV (fig.\,\ref{fig:NaISpectrum}, upper spectrum). In order to speed up the measurements, the resonant yield was taken to include all the $\gamma$-rays observed in a wide region of interest ranging from 4-8\,MeV, where the no-beam background is negligible. This region is dominated by $^{21}$Ne(p,$\gamma$)$^{22}$Na $\gamma$ rays, as was verified by a run with a HPGe detector (fig.\,\ref{fig:NaISpectrum}, lower spectrum). 

Now, for a given target gas pressure $p_{\rm T}$ and beam intensity $I$, the proton beam energy was changed in 0.5-3\,keV steps in order to populate the resonance at different positions along the beam axis. The observed resonant yield has a maximum when the resonance is populated in front of the detector. By determining the beam energy $E_{\rm p, max}$ for which a maximum is observed, the experimental energy loss $\Delta E_{\rm p}^{\rm exp}(p_{\rm T},I)$ in the gas target up to the position of maximum $\gamma$-detection efficiency is determined by subtracting the well-known resonance energy $E_{\rm p, res}$ \cite{Becker92-ZPA}:
\begin{equation}
\Delta E_{\rm p}^{\rm exp}(p_{\rm T},I) = E_{\rm p, max} - E_{\rm p, res}
\end{equation}
The experimental gas target density $n$ is then given by 
\begin{equation}\label{eq:RelDensity}
\frac{n}{n_0} = \frac{\Delta E_{\rm p}^{\rm exp}(p_{\rm T},I)}{\Delta E_{\rm p}^{\rm exp}(p_{\rm T},0)}
\end{equation}
where $n_0$ is the number density of the target gas at temperature $T$ without beam heating correction, and \linebreak $\Delta E_{\rm p}^{\rm exp}(p_{\rm T},0)$ is the proton beam energy loss obtained by extrapolating the $\Delta E_{\rm p}^{\rm exp}(p_{\rm T},I)$ curve to $I$ = 0\,$\mu$A. This definition avoids the uncertainty due to the localization of the NaI detector efficiency maximum. 

It is known from previous work \cite{Goerres80-NIM,Bemmerer06-NPA,Marta06-NIMA} that the beam heating effect is proportional to the power dissipated by the beam inside the gas per unit length, which, in turn, is given by:
\begin{equation} \label{eq:dWdx}
\frac{{\rm d}W}{{\rm d}x} = \frac{dE}{d(nx)}nI
\end{equation}
where $dE/d(nx)$ is the energy loss of protons in neon gas. For the present purposes, the value for $dE/d(nx)$ is taken from SRIM \cite{Ziegler10-NIMB}.

\begin{figure}[]
\includegraphics[width=\columnwidth]{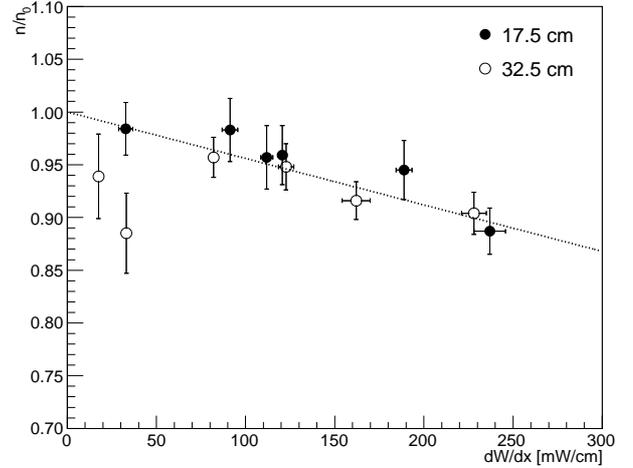}
\caption{\label{fig:BeamHeating} Ratio of observed and expected energy loss $n/n_0$ (eq.\,\ref{eq:RelDensity}) as a function of the power deposited by the ion beam per unit length ${\rm d}W/{\rm d}x$ (eq.\,\ref{eq:dWdx}), for two different positions of the NaI detector along the beam axis.}
\end{figure}

Therefore, the relative density $n/n_0$ obtained by eq.\, (\ref{eq:RelDensity}) is plotted as a function of ${\rm d}W/{\rm d}x$ (fig.\,\ref{fig:BeamHeating}). The data have been obtained by varying the gas target pressure ($p_{\rm T}$ = 0.5 , 2.5, and 4\,mbar) and beam intensity ($I$ = 30-260\,$\mu$A). Two different positions of the NaI detector along the beam axis have been used. The uncertainty for $n/n_0$ (fig.\,\ref{fig:BeamHeating}) is dominated by the uncertainty in the above described experimental determination of $\Delta E_{\rm p}^{\rm exp}(p_{\rm T},0)$, which leads to larger error bars for low pressures and low dissipated power. The error bars for ${\rm d}W/{\rm d}x$ are given by variations in the beam intensity from run to run.

The present beam heating data (fig.\,\ref{fig:BeamHeating}) clearly confirm the linear behaviour expected from previous work using other target gases and beams \cite{Goerres80-NIM,Bemmerer06-NPA,Marta06-NIMA}. The target density can therefore be described by the following relation:
\begin{eqnarray}
\frac{n}{n_0} & = & 1 - \alpha \times \frac{{\rm d}W}{{\rm d}x} \left[\frac{\rm mW}{\rm cm}\right] \\
 & = & 1 - (0.44\pm0.05) \cdot 10^{-3} \times \frac{{\rm d}W}{{\rm d}x} \left[\frac{\rm mW}{\rm cm}\right] \label{eq:BeamHeatFinal}
\end{eqnarray}

In previous static gas target experiments using neon gas, beam heating corrections were neglected \cite{Goerres82-NPA,Goerres83-NPA,Jaeger01-PRL}. The present value for $\alpha$, which was measured with beam and target diameters typical of low-energy astrophysics experiments, may serve to precisely account for this effect in future experiments for example on the $^{22}$Ne($\alpha$,n)$^{25}$Mg neutron source reaction. 

The beam heating parameter $\alpha$ = $(0.44\pm0.05) \times 10^{-3}$ determined here for proton beam in neon gas has to be compared to $\alpha$ = 0.5$\times 10^{-3}$ for proton beam in nitrogen gas \cite{Goerres80-NIM,Bemmerer06-NPA}, and  $(0.91\pm0.19)\times 10^{-3}$ for $^4$He$^+$ beam in helium gas \cite{Marta06-NIMA}, from former studies. The differences may be explained by the different heat transport coefficients in different gases. As a result, the beam heating effect is lower than in previous similar experiments at LUNA but must still be taken into account using eq.\,(\ref{eq:BeamHeatFinal}) when determining the final target density.

\section{Summary and outlook}
\label{sec:Summary}

The feasibility of the planned study of the $^{22}$Ne(p,$\gamma$)$^{23}$Na reaction at LUNA has been evaluated using an in-beam test with the setup of a previous experiment at LUNA. The $\gamma$-ray background induced by the ion beam was studied, and the main sources of background identified and discussed. 

During the tests, the resonance at $E_{\rm p}$ = 186\,keV in the $^{22}$Ne(p,$\gamma$)$^{23}$Na reaction was observed for the first time. An experimental lower limit of $\omega \gamma$ $\geq$ 0.12\,$\times$\,10$^{-6}$\,eV has been determined for the strength of this resonance.

The final experimental setup for the planned study of this reaction has been developed and described. The density profile has been studied both with and without beam. For the first time, the strength of the beam heating effect has been determined for neon gas using the resonance scan technique. This result may prove useful for future experiments using neon gas targets also beyond the present project. Results on the dedicated $^{22}$Ne(p,$\gamma$)$^{23}$Na campaign will be presented in a forthcoming publication.

\subsection*{Acknowledgments}

Financial support by INFN, by DFG (BE 4100/2-1), by the Helmholtz Association Nuclear Astrophysics Virtual Institute (NAVI, HGF VH-VI-417), and by OTKA \linebreak (K101328 and K108459) is gratefully acknowledged.


\begin{thebibliography}{10}

\bibitem{Carretta09-AA}
E.~{Carretta} {\em et~al.},
\newblock Astron.~Astrophys. {\bf 505}, 117 (2009).

\bibitem{Izzard07-AA}
R.~G. {Izzard}, M.~{Lugaro}, A.~I. {Karakas}, C.~{Iliadis}, and M.~{van Raai},
\newblock Astron.~Astrophys. {\bf 466}, 641 (2007).

\bibitem{Decressin07-AA}
T.~{Decressin}, G.~{Meynet}, C.~{Charbonnel}, N.~{Prantzos}, and
  S.~{Ekstr{\"o}m},
\newblock Astron.~Astrophys. {\bf 464}, 1029 (2007).

\bibitem{Doherty14-MNRAS3}
C.~L. {Doherty} {\em et~al.},
\newblock Monthly Notices Royal Astronom. Soc. {\bf 441}, 582 (2014).

\bibitem{Sallaska10-PRL}
A.~L. {Sallaska} {\em et~al.},
\newblock Phys.~Rev.~Lett. {\bf 105}, 152501 (2010).

\bibitem{Sallaska11-PRC}
A.~L. {Sallaska} {\em et~al.},
\newblock Phys.~Rev.~C {\bf 83}, 034611 (2011).

\bibitem{Lundmark21-PASP}
K.~{Lundmark},
\newblock Publ. Astron. Soc. Pacific {\bf 33}, 225 (1921).

\bibitem{Jose98-ApJ}
J.~Jos{\'e} and M.~Hernanz,
\newblock Astrophys.~J. {\bf 494}, 680 (1998).

\bibitem{Iliadis02-ApJSS}
C.~Iliadis, A.~Champagne, J.~Jos{\'e}, S.~Starrfield, and P.~Tupper,
\newblock Astrophys.~J.~Suppl.~Ser. {\bf 142}, 105 (2002).

\bibitem{Jose04-ApJ}
J.~Jos\'e, M.~Hernanz, S.~Amari, K.~Lodders, and E.~Zinner,
\newblock Astrophys.~J. {\bf 612}, 414 (2004).

\bibitem{Rolfs88-Book}
C.~Rolfs and W.~Rodney,
\newblock {\em Cauldrons in the Cosmos} (University of Chicago Press, Chicago,
  1988).

\bibitem{Longland10-PRC}
R.~{Longland} {\em et~al.},
\newblock Phys.~Rev.~C {\bf 81}, 055804 (2010).

\bibitem{Keinonen77-PRC}
J.~{Keinonen}, M.~{Riihonen}, and A.~{Anttila},
\newblock Phys.~Rev.~C {\bf 15}, 579 (1977).

\bibitem{Goerres82-NPA}
J.~{G\"orres}, C.~{Rolfs}, P.~{Schmalbrock}, H.~P. {Trautvetter}, and
  J.~{Keinonen},
\newblock Nucl.~Phys.~A {\bf 385}, 57 (1982).

\bibitem{Powers71-PRC}
J.~R. {Powers}, H.~T. {Fortune}, R.~{Middleton}, and O.~{Hansen},
\newblock Phys.~Rev.~C {\bf 4}, 2030 (1971).

\bibitem{Hale01-PRC}
S.~E. {Hale} {\em et~al.},
\newblock Phys.~Rev.~C {\bf 65}, 015801 (2001).

\bibitem{NACRE99-NPA}
C.~{Angulo} {\em et~al.},
\newblock Nucl.~Phys.~A {\bf 656}, 3 (1999).

\bibitem{Iliadis10-NPA841_251}
C.~{Iliadis}, R.~{Longland}, A.~E. {Champagne}, and A.~{Coc},
\newblock Nucl.~Phys.~A {\bf 841}, 251 (2010).

\bibitem{Costantini09-RPP}
H.~{Costantini} {\em et~al.},
\newblock Rep.~Prog.~Phys. {\bf 72}, 086301 (2009).

\bibitem{Broggini10-ARNPS}
C.~Broggini, D.~Bemmerer, A.~Guglielmetti, and R.~Menegazzo,
\newblock Annu. Rev. Nucl. Part. Sci. {\bf 60}, 53 (2010).

\bibitem{Scott12-PRL}
D.~A. {Scott} {\em et~al.},
\newblock Phys.~Rev.~Lett. {\bf 109}, 202501 (2012).

\bibitem{Anders14-PRL}
M.~Anders {\em et~al.},
\newblock Phys. Rev. Lett. {\bf 113}, 042501 (2014).

\bibitem{Caciolli09-EPJA}
A.~{Caciolli} {\em et~al.},
\newblock Eur.~Phys.~J.~A {\bf 39}, 179 (2009).

\bibitem{Bemmerer05-EPJA}
D.~Bemmerer {\em et~al.},
\newblock Eur.~Phys.~J.~A {\bf 24}, 313 (2005).

\bibitem{Szucs10-EPJA}
T.~{Sz{\"u}cs} {\em et~al.},
\newblock Eur.~Phys.~J.~A {\bf 44}, 513 (2010).

\bibitem{Anders13-EPJA}
M.~{Anders} {\em et~al.},
\newblock Eur.~Phys.~J.~A {\bf 49}, 28 (2013).

\bibitem{Firestone07-NDS23}
R.~B. {Firestone},
\newblock Nucl.~Data~Sheets {\bf 108}, 1 (2007).

\bibitem{Casella02-NIMA}
C.~{Casella} {\em et~al.},
\newblock Nucl.~Inst.~Meth.~A {\bf 489}, 160 (2002).

\bibitem{Formicola03-NIMA}
A.~{Formicola} {\em et~al.},
\newblock Nucl.~Inst.~Meth.~A {\bf 507}, 609 (2003).

\bibitem{Ziegler10-NIMB}
J.~F. Ziegler, M.~D. Ziegler, and J.~P. Biersack,
\newblock Nucl.~Inst.~Meth.~B {\bf 268}, 1818 (2010).

\bibitem{Agostinelli03-NIMA}
S.~Agostinelli {\em et~al.},
\newblock Nucl.~Inst.~Meth.~A {\bf 506}, 250 (2003).

\bibitem{Menzel12-Diplom}
M.-L. Menzel,
\newblock {{Experimental Study of the $^{22}${Ne}(p,$\gamma$)$^{23}${Na}
  Reaction and its Implications for Astrophysical Novae}},
\newblock {Report HZDR-034 (2012) and Diploma Thesis, Technical University of
  Dresden}, 2012.

\bibitem{Bird02-Book}
R.~B. Bird, W.~E. Stewart, and E.~N. Lightfoot,
\newblock {\em Transport phenomena}, {2nd} ed. ({John Wiley \& Sons}, 2002).

\bibitem{Goerres80-NIM}
J.~{G\"orres}, K.~{Kettner}, H.~{Kr\"awinkel}, and C.~{Rolfs},
\newblock Nucl.~Inst.~Meth. {\bf 177}, 295 (1980).

\bibitem{Bemmerer06-NPA}
D.~Bemmerer {\em et~al.},
\newblock Nucl.~Phys.~A {\bf 779}, 297 (2006).

\bibitem{Marta06-NIMA}
M.~{Marta} {\em et~al.},
\newblock Nucl.~Inst.~Meth.~A {\bf 569}, 727 (2006).

\bibitem{Becker92-ZPA}
H.~W. {Becker} {\em et~al.},
\newblock Z.~Phys.~A {\bf 343}, 361 (1992).

\bibitem{Goerres83-NPA}
J.~{G{\"o}rres} {\em et~al.},
\newblock Nucl.~Phys.~A {\bf 408}, 372 (1983).

\bibitem{Jaeger01-PRL}
M.~{Jaeger} {\em et~al.},
\newblock Phys.~Rev.~Lett. {\bf 87}, 202501 (2001).

\end{thebibliography}
\end{document}